\documentclass[letterpaper, 10 pt, final]{support/IEEETran}

\usepackage{amsmath}

\usepackage{bm}
\usepackage{cite}
\usepackage{tikz}
\usepackage{array}
\usepackage{float}
\usepackage{amsmath}
\usepackage{amssymb}
\usepackage{comment}
\usepackage{calrsfs}
\usepackage{amssymb}
\usepackage{makecell}
\usepackage{textcomp}
\usepackage{graphicx}
\usepackage{epstopdf}
\usepackage{amsfonts}
\usepackage{calligra}
\usepackage{mathrsfs}
\usepackage{mathtools}
\usepackage{subcaption}
\usepackage{color, soul}
\usepackage{nicefrac,xfrac}
\usepackage{lipsum,adjustbox}
\usepackage{tabularx,colortbl}
\usepackage{letltxmacro,xparse}
\usepackage[hidelinks]{hyperref}
\usepackage[hang,flushmargin]{footmisc}

\usepackage[T1]{fontenc}
\usepackage[ttdefault=true]{AnonymousPro}

\usetikzlibrary{shapes}
\usetikzlibrary{shapes.geometric, arrows, fit}

\graphicspath{{figures/}}
\usetikzlibrary{arrows}
\definecolor{yellow}{RGB}{255,255,0}
\definecolor{red}{RGB}{255,91,51}
\definecolor{blue}{RGB}{8,191,223}
\definecolor{green}{RGB}{54,203,30}
\definecolor{grey}{RGB}{170,170,170}
\definecolor{black}{RGB}{0,0,0}
\definecolor{orange}{RGB}{255,140,0}
\definecolor{yellow}{RGB}{225,249,27}
\tikzset{VertexStyle/.style = {shape = rectangle,fill = gray}}
\newcommand\blfootnote[1]{%
 \begingroup
 \renewcommand\thefootnote{}\footnote{#1}%
 \addtocounter{footnote}{-1}%
 \endgroup}

\def\footnoterule{\relax%
 \kern15pt
 \hbox to \columnwidth{\hfill\vrule width 1\columnwidth height 0.6pt\hfill}
 \kern4.6pt}

\LetLtxMacro\oldhref\href
\RenewDocumentCommand{\href}{o m m}{%
 \IfValueTF{#1}
 {\oldhref[#1]{#2}{\bfseries #3}}
 {\oldhref{#2}{\bfseries #3}}%
}

\DeclareMathOperator*{\argmin}{argmin}
\DeclareMathOperator*{\argmax}{argmax}

\newcommand{\norm}[1]{\left\lVert#1\right\rVert}

\newcommand{\apre}{{\textit{a priori }}}
\newcommand{\apost}{{\textit{a posteriori }}}

\tikzstyle{arrow} = [thick,->,>=stealth, line width=1pt]
\tikzstyle{circ} = [circle, minimum width=1cm, minimum height=1cm, text centered, draw=black, fill=blue!30]
\tikzstyle{circ_open} = [circle, minimum width=1cm, minimum height=1cm, text centered, draw=black, fill=red!60]
\tikzstyle{decision} = [diamond, minimum width=3cm, minimum height=1cm, text centered, draw=black, fill=orange!50,font=\bfseries]
\tikzstyle{process} = [rectangle, minimum width=4.0cm, minimum height=1cm, text centered, draw=black, fill=blue!60,font=\bfseries]
\tikzstyle{startstop} = [rectangle, rounded corners, minimum width=3.0cm, minimum height=1cm,text centered, draw=black, fill=green!50,font=\bfseries]

\newcommand{\asymcloud}[2][.1]{%
\begin{scope}[#2]
\pgftransformscale{#1}%
\pgfpathmoveto{\pgfpoint{261 pt}{115 pt}} 
  \pgfpathcurveto{\pgfqpoint{70 pt}{107 pt}}
                 {\pgfqpoint{137 pt}{291 pt}}
                 {\pgfqpoint{260 pt}{273 pt}} 
  \pgfpathcurveto{\pgfqpoint{78 pt}{382 pt}}
                 {\pgfqpoint{381 pt}{445 pt}}
                 {\pgfqpoint{412 pt}{410 pt}}
  \pgfpathcurveto{\pgfqpoint{577 pt}{587 pt}}
                 {\pgfqpoint{698 pt}{488 pt}}
                 {\pgfqpoint{685 pt}{366 pt}}
  \pgfpathcurveto{\pgfqpoint{840 pt}{192 pt}}
                 {\pgfqpoint{610 pt}{157 pt}}
                 {\pgfqpoint{610 pt}{157 pt}}
  \pgfpathcurveto{\pgfqpoint{531 pt}{39 pt}}
                 {\pgfqpoint{298 pt}{51 pt}}
                 {\pgfqpoint{261 pt}{115 pt}}
\pgfusepath{fill,stroke}         
\end{scope}}

\usepackage{acronym}

    \acrodef{WVU}{West Virginia University}
    \acrodef{MAE}{Mechanical and Aerospace Engineering}

    \acrodef{IMU}{inertial measurement unit}
    \acrodef{INS}{inertial navigation system}
    \acrodef{GPS}{global positioning system}
    \acrodef{GNSS}{global navigation satellite system}
    \acrodef{LiDAR}{light detection And ranging}
    
    \acrodef{PLL}{phase lock loop}
    \acrodef{DLL}{delay lock loop}
    \acrodef{IQ}{in-phase and quadrature}
    \acrodef{SDR}{software defined radio}
    \acrodef{RINEX}{receiver independent exchange format}
    \acrodef{RTK}{real time kinematic}
    \acrodef{PPP}{precise point positioning}
    
    \acrodef{SQM}{signal quality metric}
    \acrodefplural{SQM}[SQMs]{signal quality metrics}
    \acrodef{SS}{signal strength}
    \acrodef{EL}{elevation angle}
    \acrodef{AZ}{azimuth angle}

    \acrodef{KF}{Kalman filter}
    \acrodef{EKF}{extended Kalman filter}
    \acrodef{iSAM2}{incremental smoothing and mapping}
    \acrodef{GMM}{Gaussian mixture model}
    \acrodef{LS}{least squares}
    \acrodef{NLLS}{nonlinear least squares}
    \acrodef{SLAM}{simultaneous localization and mapping}
    \acrodef{MAP}{maximum a posteriori}
    \acrodef{LARS}{least angel regression}
    
    \acrodef{m-estimator}{maximum likelihood type estimator}
    \acrodefplural{m-estimator}[m-estimators]{maximum likelihood type estimators}
    
    \acrodef{IRLS}{iteratively re-weighted least squares}
    \acrodef{DCS}{dynamic covariance scaling}
    \acrodef{MM}{max-mixtures}
    \acrodef{BCE}{batch covariance estimation}
    \acrodef{ICE}{incremental covariance estimation}
    \acrodef{BCE-AD}{batch covariance estimation over an augmented data-space}
    \acrodef{RANSAC}{random sample consensus}
    \acrodef{RRR}{realizing, reversing, recovering}
    \acrodef{RAIM}{receiver autonomous integrity monitoring}

    \acrodef{MAP}{maximum a posteriori}
    \acrodef{MLE}{maximum likelihood estimate}

    \acrodef{RSOS}{residual-sum-of-squares}
    \acrodef{HRSOS}{horizontal residual-sum-of-squares}
    \acrodef{TRSOS}{total residual-sum-of-squares}
    
\acrodef{NN}{nearest neighbor}
\acrodef{VDP}{variational Dirichlet process}

\begin{document}
    \markboth{Submitted to IEEE Robotics and Automation Letters}%
    {Watson \MakeLowercase{\textit{et al.}}: Robust Incremental State Estimation through Covariance Adaptation}

\title{Robust Incremental State Estimation through Covariance Adaptation}

\author{Ryan M. Watson$^{1}$, Jason N. Gross$^{1}$, Clark N. Taylor$^{2}$, and Robert C. Leishman$^{2}$
\thanks{$^{1}$Department of Mechanical and Aerospace Engineering, West Virginia University, Morgantown, WV}
\thanks{$^{2}$Autonomy and Navigation Technology Center, Air Force Institute of Technology}
}
    \maketitle
    
    \begin{abstract}

Recent advances in the fields of robotics and automation have spurred significant interest in robust state estimation. To enable robust state estimation, several methodologies have been proposed. One such technique, which has shown promising performance, is the concept of iteratively estimating a Gaussian Mixture Model (GMM), based upon the state estimation residuals, to characterize the measurement uncertainty model. Through this iterative process, the measurement uncertainty model is more accurately characterized, which enables robust state estimation through the appropriate de-weighting of erroneous observations. This approach, however, has traditionally required a batch estimation framework to enable the estimation of the measurement uncertainty model, which is not advantageous to robotic applications. In this paper, we propose an efficient, incremental  extension to the measurement uncertainty model estimation paradigm. The incremental covariance estimation (ICE) approach, as detailed within this paper, is evaluated on several collected data sets, where it is shown to provide a significant increase in localization accuracy when compared to other state-of-the-art robust, incremental estimation algorithms. 

\end{abstract}
    \section{Introduction}

\IEEEPARstart{T}{he} ability to infer information about the system and the operating environment is one of the key components enabling many robotic applications. To equip robotic platforms with this capability, several state estimation frameworks \cite{simon2006optimal} have been developed (e.g., the Kalman filter \cite{kalman1960new}, or the particle filter \cite{thrun2005probabilistic}).

The traditional state estimation methodologies perform efficiently and accurately when the collected observations adhere to the \apre models. However, in many robotic applications of interest, the observations can be degraded (e.g., \ac{GNSS} observations in an urban environment, or RGB observations in a low-light setting), which cause a deviation between the collected observations and the assumed models. When this deviation is present, the traditional state estimation schemes (i.e., estimators that utilize the $l^2\text{-norm}$ exclusively to construct the cost-function) can breakdown \cite{hampel1968contribution}.

To overcome the breakdown of traditional state estimators in data degraded scenarios, several robust estimation schemes have been developed. These robust estimation schemes reduce the effect that erroneous observations have on the estimation process by scaling the associated covariance matrix \cite{watson2019enabling}. To enable this covariance scaling in practice, several implementations have been developed (i.e., \acp{m-estimator} \cite{huberBook}, switchable constraints \cite{switchCon}, and \ac{DCS} \cite{agarwal2013robust}). 

To extend robust state estimation from the traditional uni-modal uncertainty model paradigm to a multi-modal implementation, the \ac{MM} \cite{maxmix} approach was developed. The \ac{MM} approach mitigates increased computation complexity generally assumed to accompany the incorporation of multi-modal uncertainty models by first assuming that the uncertainty model can be represented by a \ac{GMM}, then selecting the single Gaussian component from the \ac{GMM} that maximizes the likelihood of the individual observation given the current state estimate. 

This \ac{MM} approach was extended in \cite{watson2019enabling} to enable the iterative estimation of the \ac{GMM}, based upon the state estimation residuals, to characterize the measurement uncertainty model. Through this iterative process, the measurement uncertainty model is more accurately characterized, which enables robust state estimation through the appropriate de-weighting of erroneous observations. This approach, however, has traditionally required a batch \cite{watson2019enabling}, or fixed-lag \cite{pfeifer2019incrementally} estimation framework to enable the estimation of the measurement uncertainty model, which is not advantageous to most robotic applications, as incremental updates are usually required. Additionally, as we'll discuss in this paper, theses approaches are inefficient -- both respect to memory and computation -- in the estimation of the measurement uncertainty model.

Within this paper, we propose a novel extension to the measurement uncertainty model estimation paradigm. Specifically, we propose an efficient, incremental extension of the methodology. The efficiency of the approach is granted by incrementally adapting the uncertainty model with only a small subset of informative state estimation residuals (i.e., the state estimation residuals which do not adhere to the \apre model). The incremental nature of the approach is granted through recent advances within the probabilistics graphical model community (i.e., through the utilization of the \ac{iSAM2} \cite{kaess2012isam2} algorithm), in conjunction with the ability to merge \ac{GMM}'s \cite{song2005highly}. \blfootnote{All software developed to enable the evaluation presented in this study is publicly available at \href{https://github.com/wvu-navLab/ICE}{https://github.com/wvu-navLab/ICE}.}

To provide a discussion of the proposed \ac{ICE} approach, the remainder of the paper is accordingly organized. First, a brief introduction to state estimation is provided in Section \ref{sec:state_est}, with a specific emphasis being placed on the current limitations of robust state estimation. Based upon the discussion provided in Section \ref{sec:state_est}, the discussion turns to the proposed \ac{ICE} robust framework in Section \ref{sec:prop_app}. In Section \ref{sec:results}, the proposed \ac{ICE} approach is validated on several collected \ac{GNSS} data sets, where improved estimation accuracy is observed, when compared to other state-of-the-art robust state estimators. Finally, the paper terminates in Section \ref{sec:conclusion} with a brief  conclusion and discussion of future research. 

    \section{State Estimation} \label{sec:state_est}

\subsection{Batch Estimation} \label{sec:batch_est}

The problem generally termed state estimation is primarily concerned with finding the set of states $X$ (i.e., a set of parameters that describe the system of interest) that is in accordance with the set of provided information $Y$. To evaluate the level of accordance between the set of states and the provided information, it is common to utilize the conditional distribution presented in Eq. \ref{eq:l2-est} (i.e., the optimal state estimate $\hat{X}$ is the state vector that maximizes the probability of the set of states conditioned on the provided information).

\begin{align}
 \hat{X} = \argmax_X \operatorname{p}(X \ | \ Y)
 \label{eq:l2-est}
\end{align}

To enable the efficient representation of this estimation problem, the factor graph \cite{dellaert2017factor} has been extensively utilized\footnote{For a \ac{GNSS} specific application, the reader is referred to \cite{gnssBayes} where the \ac{GNSS} carrier-phase ambiguity problem was equated to loop-closures in the \ac{SLAM} formulation.}. This representation is utilized because it enables the factorization of the complex \apost distribution into the product of simplified functions, as presented in Eq. \ref{eq:fg_factorization}. Where, within Eq. \ref{eq:fg_factorization}, $\psi_n$ is a single factor within the factorization, $A_n \subseteq \{ X_1, X_2 \ldots, X_N\}$, and $B_n \subseteq \{ Y_1, Y_2 \ldots Y_M \}$. 

\begin{equation}
 \operatorname{p}(X \ | \ Y) \propto \prod_{n=1}^{N} \psi_n(A_n,B_n),
 \label{eq:fg_factorization}
\end{equation}

With the factorization of the \apost distribution, as presented in Eq. \ref{eq:fg_factorization}, the state estimation problem simplifies to the canonical \ac{LS} form \cite{dellaert2017factor}, as presented in Eq. \ref{eq:nlls_cost}, where $h_n$ is the measurement function (i.e., a function that maps the state estimate to the measurement domain) and $\norm{*}$ is the $l^2\text{-norm}$. However, it should be noted that this simplification is only true if it is assumed that all of the factors within the factorization adhere to a Gaussian model \cite{dellaert2017factor}. 

\begin{equation}
 \hat{X} = \argmin_X \sum_{n=1}^{N} \lvert \lvert \ r_n(X) \ \rvert \rvert_{\Lambda_n} \quad \text{s.t.} \quad r_n(X) \triangleq y_n - h_n(X),
 \label{eq:nlls_cost}
\end{equation}

In general, for the non-linear case, there is no direct solution to the problem presented in Eq. \ref{eq:nlls_cost}. Thus, an incremental methodology of the form $X_t = X_{t-1} + \hat{\Delta}_X$ must be employed. To find an incremental update to the state estimate, it is common to linearize the measurement function about the current state estimation, as presented in Eq. \ref{eq:linear-l2-1}. The linearized representation of the estimation problem presented in Eq. \ref{eq:linear-l2-1} can be simplified by pulling the covariance matrix inside the norm, as presented in Eq. \ref{eq:linear-l2-2}, where $a_n$ and $b_n$ are the whitened measurement Jacobian and state estimation residual vectors (i.e., $a_n \triangleq \Lambda_n^{-\nicefrac{1}{2}} \frac{\partial h_n(X_{t-1})}{\partial X}, \ \text{and} \ b_n \triangleq \Lambda_n^{-\nicefrac{1}{2}} r_n$), respectively. 

\begin{align}
     \hat{\Delta}_X &= \argmin_{\Delta_X} \sum_{n=1}^N \norm{  \frac{\partial h_n(X_{t-1})}{\partial X} \Delta_X - r_n}_{\Lambda_n} \label{eq:linear-l2-1}\\ 
    &=  \argmin_{\Delta_X} \sum_{n=1}^N \norm{ a_n \Delta_X - b_n}, \label{eq:linear-l2-2}
\end{align}

The cost function presented in Eq. \ref{eq:linear-l2-2}, can be more compactly defined as presented within Eq. \ref{eq:l2-stacked}, where the matrices $A$ and $B$ are defined by stacking vertically their respective whitened components (i.e., $A$ is a matrix formed by vertically stacking the set $\{a_1, \hdots, a_N\}$, and $B$ is a matrix formed by vertically stacking the set $\{b_1, \hdots, b_N\}$).

\begin{equation}
    \hat{\Delta}_X = \argmin_{\Delta_X} \norm{A \Delta_X - B}, 
    \label{eq:l2-stacked}
\end{equation}

To solve the system presented in Eq. \ref{eq:l2-stacked}, it is common to utilize a matrix factorization of the measurement Jacobain matrix\footnote{To make a connection back to the graphical model (i.e., the factor graph), it was shown in \cite{kaess2012isam2} that variable elimination \cite{blair1993introduction} on the factor graph (i.e., converting a factor graph to a Bayes net) is equivalent to QR-decomposition.} \cite{dellaert2006square}. For this discussion, the QR-decomposition \cite{bierman2006factorization} is utilized, which provides a factorization as presented in Eq. \ref{eq:qr}, where $Q \in \mathbb{R}^{N \times N}$ is an orthogonal matrix and $R \in \mathbb{R}^{M\times M}$ is an upper-triangular matrix.

\begin{equation}
A = Q\begin{bmatrix}R\\0\end{bmatrix},
\label{eq:qr}
\end{equation}

Utilizing the factorization presented in Eq. \ref{eq:qr}, the cost function presented in Eq. \ref{eq:l2-stacked} can equivalently\footnote{The cost functions presented in Eq. \ref{eq:l2-stacked} and Eq. \ref{eq:qr_ls_1} are equivalent due to the orthogonality of the matrix $Q$ (i.e., $\norm{Qv} = \norm{v}$ given that $Q$ is orthogonal).} be expressed as provided in Eq. \ref{eq:qr_ls_1}, which simplifies to the expression provided in Eq. \ref{eq:qr_ls_2}. The expression provided in Eq. \ref{eq:qr_ls_2} is computational efficient due to the upper triangular nature of the matrix $R$ (i.e., the system $R\hat{\Delta}_X = c$ can simply be solved via back substitution).

\begin{align}
    \hat{\Delta}_X &= \argmin_{\Delta_X} \norm{Q^T (A\Delta_X - B)} \label{eq:qr_ls_1}\\
     &= \argmin_{\Delta_X} \norm{R\Delta_X - c} + \norm{d} \label{eq:qr_ls_2} \quad \text{s.t.} \quad Q^{T}B \triangleq \begin{bmatrix} c \\ b \end{bmatrix}
\end{align}

\subsection{Incremental Estimation} \label{sec:inc_update}

The estimation framework discussed in Section \ref{sec:batch_est} provides an efficient and numerically stable solution when all of the information is provided beforehand. However, for many applications, the information is provided incrementally. When this is the case, the estimation framework discussed previously is inefficient due to the need to recompute the QR-decomposition of the entire measurement Jacobian matrix every time a new information is provided.

To overcome this computation limitation, the concept of incrementally updating the QR-decomposition was studied within \cite{kaess2007fast}. Within \cite{kaess2007fast}, they enabled the incremental updating of the matrix factorization by first augmenting the previous factorization (i.e., incorporating new rows in the $R$ and c matrices), then, restoring the upper triangular form of the factorization through the utilization of Givens rotations\footnote{See section 5.1.8 of \cite{golub13} for a thorough review of Givens rotations with applications to \ac{LS}.}. 

The approach proposed within \cite{kaess2007fast} does have one key limitation, which is the requirement to conduct periodic batch re-computation of the QR-decomposition for the entire measurement Jacobian matrix to enable variable re-ordering. This batch re-computation is utilized to maintain the sparsity of the upper-triangular system. To mitigate this batch re-computation the Bayes tree \cite{kaess2010bayes} was introduced. This directed graphical model directly represents the square root information matrix (i.e., the matrix $R$ in Eq. \ref{eq:qr_ls_2}) and can be easily computed from the associated factor graph in a two-step process, as detailed in \cite{kaess2012isam2}. Due to the structure of the Bayes tree graphical model, this methodology removes the requirement to re-factor the entire system when new information is added. Instead, only the affected section of the Bayes tree is re-factored, as detailed within \cite{kaess2012isam2}. This approach to state estimation is title \ac{iSAM2}, and is the approach utilized within this study. 

\subsection{Robust Estimation}

Utilizing the \ac{iSAM2} approach provides an efficient estimation framework when the provided information adheres to the \apre models. However, when the provided information does not adhere to the \apre models, the estimator can breakdown \cite{hampel1968contribution}. This property is not exclusive to the \ac{iSAM2} framework, instead, it is a fundamental property of any estimation framework that exclusively utilizes the $l^2\text{-norm}$ to construct it's cost function.

To overcome this limitation, several robust estimation frameworks have been proposed (e.g., \acp{m-estimator} \cite{huberBook}, switchable constraints \cite{switchCon}, and \ac{MM} \cite{maxmix}). Linking all of these estimation frameworks is the concept of enabling robust estimation through appropriately weighting (i.e., scaling the assumed covariance model) the contribution of each information source based upon the level of adherence between the information and the \apre model. To implement this concept, the  \ac{IRLS} formation \cite{zhang1997parameter}, as provided in Eq. \ref{eq:irls}, can be utilized, where the weighting function $w(*)$ is dependent upon the utilized robust estimation framework (i.e., \ac{DCS} \cite{agarwal2013robust}).

\begin{equation}
 \hat{X} = \argmin_X \sum_{n=1}^{N} w_n(e_{n}) \ e_n \quad \text{s.t.} \quad e_n \triangleq \norm{r_n(X)}_{\Lambda_n}
 \label{eq:irls}
\end{equation}

To extend robust state estimation from the traditional uni-modal uncertainty model paradigm to a multi-modal implementation, the \ac{MM} \cite{maxmix} approach was developed. The \ac{MM} approach mitigates increased computation complexity generally assumed to accompany the incorporation of multi-modal uncertainty models by first assuming that the uncertainty model can be represented by a \ac{GMM}, then selecting the single Gaussian component from the \ac{GMM} that maximizes the likelihood of the individual observation given the current state estimate.

The \ac{MM} approach was extended within the \ac{BCE} framework \cite{watson2018robust, watson2019enabling} to enable the estimation of the multi-modal covariance models during optimization. The \ac{BCE} approach enables the estimation of the multi-modal covariance model through the utilization of variational clustering \cite{varInf} on the current set of state estimation residuals. The \ac{BCE} approach provided promising results with the with the primary limitation being the batch estimation nature of the framework. To overcome this computational limitation, an extension to the \ac{BCE} approach, as described within section \ref{sec:prop_app}, which enables efficient incremental updating while maintaining the robust characteristics, is proposed within this paper.

\section{Proposed Approach} \label{sec:prop_app}

To facilitate a discussion of the proposed \ac{ICE} framework the assumed data model is first explained. Then, a method for incremental measurement uncertainty model adaptation is presented. Finally, pull the previously mentioned topics together, the discussion concludes with an overview of the proposed \ac{ICE} framework.

\subsection{Data Model} \label{sec:data_model}

As calculated by the estimator, a set of state estimation residuals $\mathbf{R} = \{ r_1, r_2, \ldots, r_N \ | \ r_n \triangleq y_n - h_n(X)\}$ is provided. The set of state estimation residuals can be characterized by a \ac{GMM}, which, for this work, will act as the measurement uncertainty model, $\text{GMM}_g$. As proposed within \cite{maxmix}, with the intent to minimize the computation complexity of the optimization problem, the \ac{GMM} can be reduced to selecting the most likely component from the mixture model to approximately characterize each observation, as depicted in Eq. \ref{eq:max_mix} where $\mu_m$ is the components mean and $\Lambda_m$ is the components covariance. 

\begin{equation}
    r_n \sim \max_m w_m \mathcal{N}(r_n \ | \ \theta_m) \quad \text{s.t.} \quad \theta_m = \{ \mu_m, \Lambda_m\}
    \label{eq:max_mix}
\end{equation}

For this work, it is additionally assumed that the set of residuals, $\mathbf{R}$, can be partitioned into two distinct groups. The first group is the set of all residuals which sufficiently adhere to the \apre covariance model (i.e., do not deviate sufficiently from the most likely component within $\text{GMM}_g$), which will be indicated by the set $\mathbf{R_{I}}$. While, the second group is the set residuals which do not sufficiently adhere to the \apre covariance model, which will be indicated by  the set $\mathbf{R_{O}}$.

To quantify the level of adherence to the \apre uncertainty model, the $\operatorname{z-test}$, as provided in Eq. \ref{eq:z-test}, is employed. Within Eq. \ref{eq:z-test} $\mu$, and $\sigma$ are the mean and standard deviation of the most likely component from $\text{GMM}_g$ for the state estimation residual $r_n$. Utilizing the $\operatorname{z-test}$ as a metric to quantify the level of agreement between the set of state estimation residual and the \apre uncertainty model, we can more concretely define the two groupings as, $\mathbf{R_{I}} = \{ r \ | \ r \in \mathbf{R}, \ Z(r, \phi) < T_r\}$\footnote{$T_r$ is a user defined parameter that encodes the acceptable amount an observation can deviation from the \apre model in terms of multiples of the standard deviation.} and $\mathbf{R_{O}} = \{ r \ | \ r \in \mathbf{R}, \ r \notin \mathbf{R_{I}} \}$.

\begin{equation}
    Z(r_n,\phi) = \frac{r_n - \mu}{\sigma} \quad \text{s.t.} \quad \phi \triangleq \{\mu, \sigma \}
    \label{eq:z-test}
\end{equation}

\subsection{Uncertainty Model Adaptation} \label{sec:model_update}

By definition, the set $\mathbf{R_O}$ is not accurately characterized by $\text{GMM}_g$ thus, it is desired to adapt the uncertainty model to more accurately represent the new observations. To enable the adaptation of the uncertainty model, a two step procedure is utilized. This procedure starts by estimating a new \ac{GMM}, which will be indicated by $\text{GMM}_n$, based solely on the set $\mathbf{R_O}$. Then, $\text{GMM}_n$ is merged into the prior model (i.e., $\text{GMM}_g$) to provide a more accurate characterization the measurement uncertainty model. This procedure is elaborated upon in Section \ref{sec:clustering} and Section \ref{sec:merging}, respectively.

\subsubsection{Variational Clustering} \label{sec:clustering}

To estimate $\text{GMM}_n$, the set of model parameters which maximizes the $\operatorname{log \ marginal \ likelihood}$, as depicted in Eq. \ref{eq:model_fit}, must be calculated. In Eq. \ref{eq:model_fit}, $\boldsymbol{\theta}$ is the set of mean vectors and covariance matrices which define the new \ac{GMM}, and $\mathbf{Z}$ is an assignment variable (i.e., the variable $\mathbf{Z}$ assigns each $r\in\mathbf{R_O}$ to a specific component within the model).

\begin{equation}
 \operatorname{log}\operatorname{p}(\mathbf{R_O}) = \operatorname{log} \int \operatorname{p}(\mathbf{R_O}, \boldsymbol{\theta}, \mathbf{Z}) d\mathbf{Z} d\boldsymbol{\theta}
 \label{eq:model_fit}
\end{equation}

In general, the integral presented in Eq. \ref{eq:model_fit} is computational intractable \cite{beal2003variational}. Thus, a method of approximate integration must be implemented. For this work, the variational inference\footnote{To enable the implementation of the \ac{ICE} approach in software, the \href{https://github.com/dsteinberg/libcluster}{libcluster} \cite{steinberg2013unsupervised} software library was utilized.} \cite{beal2003variational, steinberg2013unsupervised} approach is utilized primarily due this class of algorithms run-time performance when compared to sampling based approaches (i.e., Monte Carlo methods \cite{doucet2005monte}).

\subsubsection{Efficient GMM Merging} \label{sec:merging}

To enable the second step of the measurement uncertainty model adaptation (i.e., the merging of $\ac{GMM}_n$ into the prior model $\ac{GMM}_g$), an implementation of the algorithm presented in \cite{song2005highly} is utilized. To provide a description of the approach, let's evaluate the equivalence between $g_{n} \triangleq \{w_n, \mu_n, \Lambda_n \}\in \ac{GMM}_n$ (e.g., the first component in $\ac{GMM}_n$) and $g_{g} \triangleq \{ w_g, \mu_g, \Lambda_g\}\in \ac{GMM}_g$ (e.g., the first component in $\ac{GMM}_g$).

To test the equivalence, we will first extract the set of observations $\mathbf{R_{O,g_n}} \subseteq \mathbf{R_O}$ that correspond to set of state estimation residuals that are characterized by component $g_{n}$. Utilizing $\mathbf{R_{O,g_n}}$, it is desired to check if the set of state estimation residuals has an equivalent covariance to the hypothesis covariance model (i.e., we want to see if $\Lambda_{n} = \Lambda_{g}$, where $\Lambda_{n} = \operatorname{cov}(\mathbf{R_{O,g_n}})$ and $\Lambda_{g}$ is the hypothesis covariance from $g_{g}$).

To determine if our two \ac{GMM} components have an equivalent covariance model, we must first transform the set of observations $\mathbf{R_{O, g_n}}$ with Cholesky decomposition of our hypothesis covariance\footnote{This whitening process is conducted because the covariance test is only valid for unit covariance matrices.}. This transformation provides us with a new data set, defined as $\mathbf{Y} = \{ y = L^{-1} r \ | \ r \in \mathbf{R_{O, g_n}}, \ \Lambda_{g} = L L^T \}$.

Utilizing the transformed set of state estimation residuals $\mathbf{Y}$, the $W$-statistic \cite{ledoit2002some} can be constructed, as provided in Eq. \ref{eq:w-stat}, to test the equivalence of covariance matrices. Within Eq. \ref{eq:w-stat}, $\Lambda_y = \operatorname{cov}(\mathbf{Y}$), $m$ is the cardinality of the set $\mathbf{Y}$ (i.e., $m = \rvert \mathbf{Y} \lvert$), and $d$ is the dimension the state estimation residuals (i.e, $y_m \in \mathbb{R}^{d}$). 

\begin{equation}
    W = \frac{1}{d} Tr\big((\Lambda_y - I)^2\big) - \frac{d}{m}\big(\frac{1}{d}Tr(\Lambda_y)\big)^2 + \frac{d}{m}
    \label{eq:w-stat}
\end{equation}

The $W$-statistic is known to have an asymptotic $\chi^2$ distribution with degrees of freedom $d(d+1)/2$, as depicted in Eq. \ref{eq:cov-test}. Thus, a Chi-square test with a user defined critical value is utilized to test the equivalence of covariance matrices.

\begin{equation}
    \frac{mWd}{2} \sim \chi^2_{d(d+1)/2}
    \label{eq:cov-test}
\end{equation}

To test the equivalence of mean vectors, the $T$-statistic \cite{hotelling1992generalization}, as provided in Eq. \ref{eq:t-stat}, is utilized. Within Eq. \ref{eq:t-stat}, $\mu_{n}$ is the mean of the component of $\ac{GMM}_n$, and $\mu_{g}$ is the mean vector of the component of $\ac{GMM}_g$. The $T$-statistic is utilized to test the equivalence of mean vectors because it is known to have an asymptotic $F$ distribution, as depicted in Eq. \ref{eq:mean-test}. Thus, an F-test with user defined critical value is utilized to test the equivalence of mean vectors.

\begin{equation}
    T^2 = m\norm{\mu_{n} - \mu_{g}}_{\Lambda_{y}}
    \label{eq:t-stat}
\end{equation}

\begin{equation}
    \frac{m-d}{d(m-1)}T^2 \sim F_{d,m-d}
    \label{eq:mean-test}
\end{equation} 

If both the mean and covariance of two components are found to be equivalent, then the new component $g_{n}$ is merged with the prior component $g_{g}$ to adapt the measurement uncertainty model $\ac{GMM}_g$. To adapt the measurement uncertainty model, the mean, covariance and weighting can be updated, as presented in Eqs. \ref{eq:mean_up}, \ref{eq:cov_up}, and \ref{eq:weight_up}, respectively. Within Eqs. \ref{eq:mean_up}, \ref{eq:cov_up}, and \ref{eq:weight_up}, $N$ is the total number of points which are characterized by $\ac{GMM}_g$, $M$ is the total number of points which are characterized by $\ac{GMM}_n$, and $m$ is the number of points which are characterized by component $g_n$.

\begin{equation}
    \mu = \frac{N w_{g} \mu_{g} + m \mu_{n}}{N w_{g} + m} 
    \label{eq:mean_up}
\end{equation}

\begin{equation}
    \Lambda = \frac{N w_{g} \Lambda_{g} + m \Lambda_{n}}{N w_{g} + m} + \frac{N w_{g} \mu_{g}\mu_{g}^T + m \mu_{n}\mu_n^T}{N w_g+m} - \mu\mu^T 
    \label{eq:cov_up}
\end{equation}

\begin{equation}
    w = \frac{N w_g + m}{N + M} 
    \label{eq:weight_up}
\end{equation}

If the new component $g_n$ does not match a component within $\ac{GMM}_g$, then the mean and covariance of $g_n$ is added to $\ac{GMM}_g$. When the new component is added to $\ac{GMM}_g$ the weighting vector is updating, as presented in Eq. \ref{eq:new_weight-1}, where $N$, $M$, and $m$ are as defined above. When the new component is added, the weighting for all of the remaining components in $\ac{GMM}_g$ are updated according to Eq. \ref{eq:new_weight-2}.

\begin{equation}
    w = \frac{m}{N + M}
    \label{eq:new_weight-1}
\end{equation}

\begin{equation}
    w = \frac{N w_g}{N+M}
    \label{eq:new_weight-2}
\end{equation}

Through the utilization of the mixture model merging approach developed within \cite{song2005highly}, and outlined in this section, the measurement uncertainty model can be adapted online. This adaptation is conducted without the need for storing all previous state estimation residuals (i.e, only the most recent residuals $\mathbf{R_O}$ which do not adhere to the \apre model are required), which dramatically reduces the computational and memory cost of the proposed approach.

\subsection{Algorithm Overview}

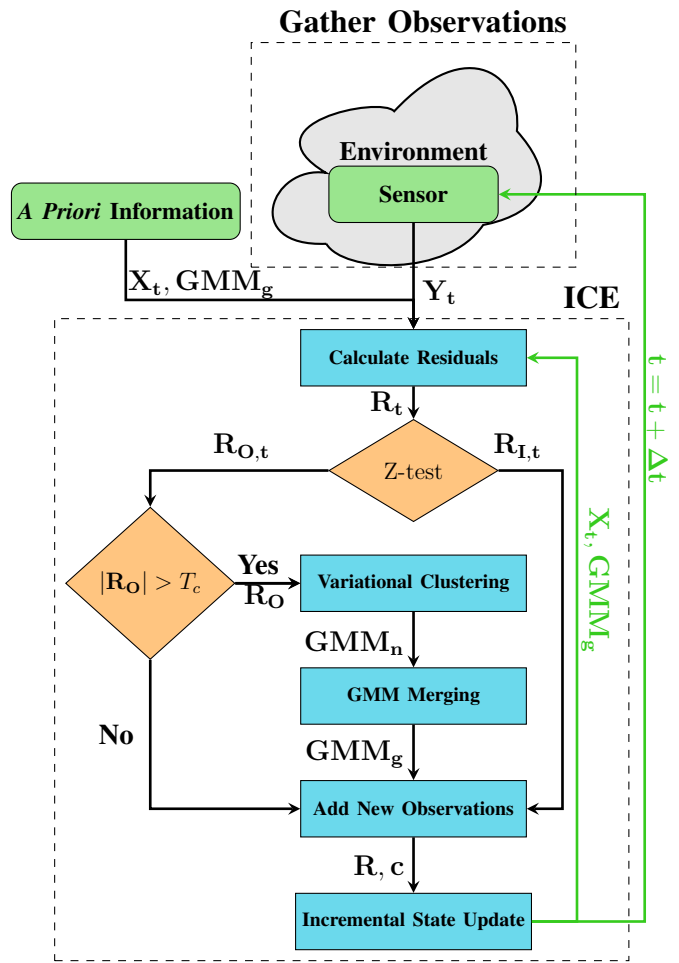
\begin{figure}
 \centering
 \begin{tikzpicture}[node distance=1.5cm]
 
   \node (pro1) at (0,0) [process, xshift=0.0cm, yshift=-0.8cm, scale=0.75] {Calculate Residuals};

  \node (cloud) [above of=pro1, xshift=0.0cm, yshift=1.2cm] {\tikz \asymcloud[0.17]{fill=gray!20,thick};};
  \node [above of=pro1, xshift=0.0cm, yshift=1.25cm] {\textbf{Environment}};
  \node (start) [startstop, above of=pro1, scale=0.75, yshift=0.90cm] {\large{Sensor}};
  
  \node (start2) [startstop,scale=0.75, above of=pro1, xshift=-5.1cm, yshift=1.1cm] {\large{\textit{A Priori} Information}};
  
  \node (dec1) [decision, below of=pro1,scale=0.75] {\large{$\operatorname{Z-test}$}};

  \node (dec2) [decision, below of=dec1, xshift=-3.5cm, scale=0.75] {\large{$|\mathbf{R_O}| > T_c$}};

  \node (pro4) [process, below of=dec1, scale=0.75] {Variational Clustering};
  \node (pro5) [process, below of=pro4, scale=0.75] {GMM Merging};

  \node (pro8) [process, below of=pro5, scale=0.75] {Add New Observations};
  \node (pro9) [process, below of=pro8, scale=0.75] {Incremental State Update};

  \draw [arrow] (pro1.south) node[above left, xshift=0.0cm, yshift=-0.5cm, scale=0.9]   {\large{$\mathbf{R_t}$}} -| (dec1.north);

  \draw [arrow] (dec1.west) node[above left, xshift=-0.65cm, scale=0.9]   {\large{$\mathbf{R_{O,t}}$}} -| (dec2.north);
  \draw [arrow] (dec1.east) node[above left, xshift=0.7cm, scale=0.9]   {\large{$\mathbf{R_{I,t}}$}} -| ++(8.5mm,0) |- (pro8.east);

  \draw [arrow] (dec2.east) node[above left, xshift=0.7cm, scale=0.9]   {\large{\textbf{Yes}}}   -- (pro4.west);
  
    \draw [arrow] (dec2.east) node[above left, xshift=0.80cm, yshift=-0.45cm, scale=0.9]   {\large{$\mathbf{R_O}$}}   -- (pro4.west);

 \draw [arrow] (dec2.south) node[above left, xshift=-0.1cm, yshift=-1.24cm, scale=0.9]   {\large{\textbf{No}}} |- ++(0,-10mm) |-  (pro8.west);

 \draw [arrow] (pro4.south) node[above left, xshift=-0.0cm, yshift=-0.7cm, scale=0.9]   {\large{$\mathbf{GMM_{n}}$}} -- (pro5.north);
 
 \draw [arrow] (pro5.south) node[above left, xshift=-0.0cm, yshift=-0.7cm, scale=0.9]   {\large{$\mathbf{GMM_{g}}$}} -- (pro8.north);

 \draw [arrow] (pro8.south) node[above left, xshift=-0.0cm, yshift=-0.7cm, scale=0.9]   {\large{$\mathbf{R, c}$}} -- (pro9.north);
   
\draw [green, arrow] (pro9.east) node (fb1) [above left, xshift=0.54cm, yshift=3.45cm, rotate=-90, scale=0.9]   {\large{$\mathbf{X_t, GMM_g}$}} -| ++(6mm,0) |- (pro1.east);

\draw [green, arrow] (pro9.east) node [above left, xshift=1.41cm, yshift=5.7cm, scale=0.86, rotate=-90]   {\large{$\mathbf{t = t + \Delta t}$}} -| ++(15mm,0) |- (start.east);

 \draw [arrow] (start.south) node[above left, xshift=0.7cm, yshift=-1.2cm, scale=0.9]   {\large{$\mathbf{Y_t}$}} -- (pro1.north);
 
  \draw [arrow] (start2.south) node[above left, xshift=2.1cm, yshift=-0.9cm, scale=0.9]   {\large{$\mathbf{X_t, GMM_{g}}$}} |- ++(0,-8mm) -| (pro1.north);

\node (db2) [draw,inner sep=4pt, dashed,fit=(pro1) (pro4) (pro5) (dec2) (pro8) (pro9) (fb1)] {};
\node[above left] at (db2.north east) { {\textbf{\large{ICE}} }};

\node (db1) [draw,inner sep=-8pt, dashed,fit=(cloud)] {};
\node[above left] at (db1.north east) { {\textbf{\large{Gather Observations}} }};

 \end{tikzpicture}
 \caption{Graphical depiction of the proposed incremental covariance estimation (ICE) algorithm. The proposed approach enables efficient, incremental, and robust state estimation through the iterative adaptation of the measurement uncertainty model, based upon the state estimation residuals that correspond to erroneous observations.}
  \label{fig:algo_ice}
\end{figure}

With the discussion provided in the previous sections, the conversation can now turn to an overview of the proposed robust estimation framework. To facilitate a discussion, a graphical overview of the \ac{ICE} framework is depicted in Fig. \ref{fig:algo_ice}.

From Fig. \ref{fig:algo_ice}, it is shown that the \ac{ICE} algorithm starts at each epoch by calculating the set of state estimation residuals $\mathbf{R_t}$ from the current set of observations $\mathbf{Y_t}$. As discussed within Section \ref{sec:data_model}, this set of state estimation residuals $\mathbf{R_t}$ can be partitioned into two distinct groups (i.e., the set of state estimation residuals which correspond to erroneous observations $\mathbf{R_{O,t}}$, and the set of state estimation residuals which correspond to observations that adhere to the \apre model $\mathbf{R_{I,t}}$) through the utilization of the $\operatorname{z-test}$.

With the set $\mathbf{R_{O,t}}$, the previous set of state estimation residuals which correspond to erroneous observations $\mathbf{R_O}$ is appended. If the length of $\mathbf{R_O}$ is greater than a user defined threshold\footnote{Several factors can affect the specific realization of this threshold (e.g., the expected dynamics of the environment, or the number of observations per epoch).} (i.e., if $\rvert \mathbf{R_O} \lvert > T_c$), the set is utilized to modify the measurement uncertainty model, as described in Section \ref{sec:model_update}. After the adaptation of the uncertainty model, the set $\mathbf{R_O}$ is cleared and the set of observations which adhere to the \apre model $\mathbf{R_{I,t}}$ are incorporated. With the incorporation of the new observations, a new state estimate is provided, following the discussion provided in Section \ref{sec:inc_update}.

If the length of set of state estimation residuals, which correspond to erroneous observations, $\mathbf{R_O}$ is less than a user defined threshold, then the uncertainty model is not adapted for the current epoch. Instead, the previous measurement uncertainty model is utilized to incorporate the new set of observations which adhere to the \apre model. With the new observations incorporated, a new state estimated is provided, as described in Section \ref{sec:inc_update}. This process is continued in an iterative fashion for as long as needed (e.g., until the data collection terminates).

    \section{Results} \label{sec:results}
\subsection{Data Collection}

\begin{figure}
 \centering
 \includegraphics[width=0.9\linewidth]{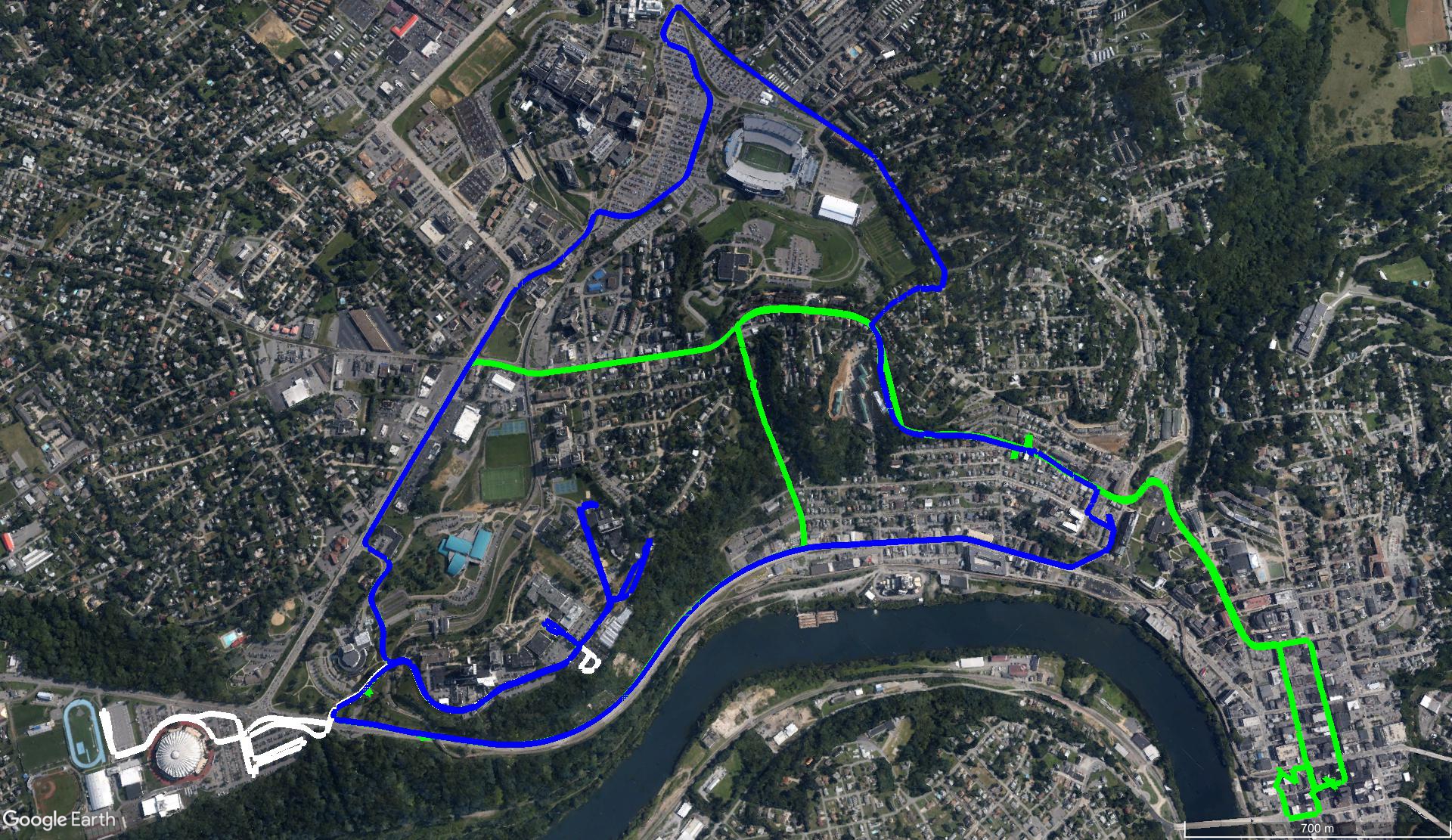}
 \caption{Ground trace for the three utilized \ac{GNSS} data sets. The white trace corresponds to data collect 1, the green trace corresponds to data collect 2, and the blue trace corresponds to data collect 3.}
 \label{fig:ground_trace}
\end{figure}

To conduct an evaluation of the proposed robust estimation framework, a collection of three kinematic \ac{GNSS} data sets is utilized. These \ac{GNSS} data sets, as can be visualized through their ground traces, which are shown in Fig. \ref{fig:ground_trace}, were made publicly available and are described within~\cite{watson2019enabling}.

For these data collects, the binary \ac{IQ} data in the L1-band was recorded. By recording the \ac{IQ} data in place of the \ac{GNSS} receiver dependent observations (i.e., the pseudorange and carrier-phase observables), the same data collect can be utilized to generate several sets of observations with varying levels of degradation after playing back through a software defined \ac{GNSS} receiver \cite{fernandez2001gnss} with different sets of tracking parameters. Specifically, the receiver dependent observations can be generated off-line by playing the \ac{IQ} data into a \ac{GNSS} receiver, where the level of degradation is varied by changing the \ac{GNSS} receiver's tracking parameters (i.e., changing the bandwidth of the \ac{PLL}, the \ac{DLL} and the correlator spacing). For a detailed discussion on the impact that the \ac{GNSS} receiver tracking parameters can have on the quality of the generated observables, the reader is referred to \cite{kaplan2005understanding,van1992theory}, which is reviewed in \cite{watson2019enabling}.

For this study, two sets of observations are generated (i.e., a low-quality and high-quality data set) for each of the data collects. The specific \ac{GNSS} receiver tracking parameters utilized to generate the low-quality and high-quality observations are provided within Table III of \cite{watson2019enabling}

\subsection{Evaluation}

Utilizing these data collects, an evaluation of the proposed methodology can be conducted. To provided a comparison for the proposed approach, three additional estimation frameworks will be utilized. The first comparison methodology is the traditional $l^2\text{-norm}$ based estimator. The second comparison methodology is the \ac{MM} approach, which has a static measurement error covariance model (i.e., a fixed two component measurement error covariance model). The final comparison methodology is the \ac{DCS} approach, where the \ac{DCS} approach is utilized because it is both a closed form version of switchable constraints and a specific implementation of an \ac{m-estimator} \cite{agarwal2013robust}. All of the utilized estimators are built upon the \ac{iSAM2} algorithm \cite{kaess2012isam2}, as implemented within the Georgia Tech Smoothing and Mapping (\href{https://gtsam.org/}{GTSAM}) library \cite{dellaert2012factor}.

\subsubsection{Localization Performance}

To start an evaluation, the localization performance of the estimation frameworks will be assessed. To enable the assessment of the localization performance, a reference ground-truth must first be established. To generate this ground-truth, a differential \ac{GNSS} solution (i.e., \ac{RTK}\footnote{This solution was realized with \href{http://www.rtklib.com/}{RTKLIB}\cite{takasu2011rtklib}, which is an open-source software package for \ac{GNSS} based localization.}) is utilized, which is known to provide centimeter level localization accuracy \cite{kaplan2005understanding}. 

With the \ac{RTK} generated reference ground-truth solution, the localization performance of the four estimation frameworks, when low-quality observations are utilized, is provided in Table \ref{table:lq_stats}\footnote{ The localization performance presented within this section is significantly improved from the batch implementation presented within \cite{watson2019enabling}. This localization performance increase is primarily due to two modifications: 1) an accurate carrier-phase cycle slip threshold was set, 2) a static position constraint is placed on the initial and final positions within this implementation. \\}. From Table \ref{table:lq_stats}, it can be seen that all three of the robust estimation frameworks provided a significant increase is localization accuracy, with respect to the median, when compared to the traditional $l^2\text{-norm}$ approach. Additionally, it should be noted that the \ac{ICE} approach provides the most accurate solution for all three data collects when low-quality observations are utilized.

\begin{table*}
 \caption{Horizontal RSOS localization error results when low fidelity receiver tracking parameters are utilized to generate the observations. The green and red cell entries correspond to the minimum and maximum statistic, respectively.}
 \begin{subtable}{0.3\linewidth}
      \centering
      \caption{Localization results for data collect 1.}
     \resizebox{1.0\linewidth}{!}{%
      \begin{tabular}{l|ccccc}
          \hline
          \cellcolor{grey!30}{(m.)}     & \cellcolor{grey!30}{$L_2$} & \cellcolor{grey!30}{DCS}   & \cellcolor{grey!30}{MM}     & \cellcolor{grey!30}{ICE}                         \\
          \hline
          \cellcolor{grey!30}{mean}   & \cellcolor{red!60}{2.51}  & 0.99 & 1.66    & \cellcolor{green!60}{0.73}  \\
          \cellcolor{grey!30}{median}   & \cellcolor{red!60}{2.57}  & 0.64 & 1.63    & \cellcolor{green!60}{0.56}  \\
          \cellcolor{grey!30}{std. dev.} & \cellcolor{red!60}{1.41}  & 0.98              & 1.05        & \cellcolor{green!60}{0.72}  \\
          \cellcolor{grey!30}{max}      & 10.78 & \cellcolor{green!60}{9.71} & 10.06   & \cellcolor{red!60}{13.19} \\
          \hline
      \end{tabular}%
      }
      \label{table:drive_1_lq}
 \end{subtable} \hfill
 \begin{subtable}{0.3\linewidth}
      \centering
      \caption{Localization results for data collect 2.}
        \resizebox{1.0\linewidth}{!}{%
      \begin{tabular}{l|ccccc}
          \hline
          \cellcolor{grey!30}{(m.)}     & \cellcolor{grey!30}{$L_2$}                    & \cellcolor{grey!30}{DCS}                          & \cellcolor{grey!30}{MM}   & \cellcolor{grey!30}{ICE}                        \\
          \hline
              \cellcolor{grey!30}{mean}   & \cellcolor{red!60}{4.00} & \cellcolor{red!60}{4.00}      & 3.12               & \cellcolor{green!60}{2.11} \\
          \cellcolor{grey!30}{median}   & \cellcolor{red!60}{2.48} & 2.08                         & 1.94               & \cellcolor{green!60}{0.93} \\
          \cellcolor{grey!30}{std. dev.} & 3.87                   & \cellcolor{red!60}{4.59} & 3.92 & \cellcolor{green!60}{2.10}                     \\
          \cellcolor{grey!30}{max}      & 29.18   & 31.05  & \cellcolor{red!60}{31.40} & \cellcolor{green!60}{23.02}                     \\
          \hline
      \end{tabular}%
      }
      \label{table:drive_2_lq}
 \end{subtable} \hfill
 \begin{subtable}{0.3\linewidth}
      \centering
      \caption{Localization results for data collect 3.}
      \resizebox{1.0\columnwidth}{!}{%
      \begin{tabular}{l|ccccc}
          \hline
          \cellcolor{grey!30}{(m.)}     & \cellcolor{grey!30}{$L_2$}                     & \cellcolor{grey!30}{DCS}                        & \cellcolor{grey!30}{MM}                     & \cellcolor{grey!30}{ICE}                       \\
          \hline
          \cellcolor{grey!30}{mean}   & \cellcolor{red!60}{4.94} & \cellcolor{green!60}{4.16}  & 4.51      & 4.35 \\
          \cellcolor{grey!30}{median}   & \cellcolor{red!60}{4.41} & 2.82        & 3.62      & \cellcolor{green!60}{1.48} \\
          \cellcolor{grey!30}{std. dev.} & \cellcolor{green!60}{2.97}       & 3.54 & 3.33   & \cellcolor{red!60}{5.23}                      \\
          \cellcolor{grey!30}{max}      & 29.53    & \cellcolor{red!60}{30.38}  & 28.30 & \cellcolor{green!60}{26.61}                      \\
          \hline
      \end{tabular}%
      }
      \label{table:drive_3_lq}
 \end{subtable}
 \label{table:lq_stats}
\end{table*}

\begin{table*}
 \caption{Horizontal RSOS localization error results when high fidelity receiver tracking parameters are utilized to generate the observations. The green and red cell entries correspond to the minimum and maximum statistic, respectively.}
 \begin{subtable}{0.3\linewidth}
      \centering
      \caption{Localization results for data collect 1.}
      \resizebox{1.0\columnwidth}{!}{%
      \begin{tabular}{l|ccccc}
           \hline
           \cellcolor{grey!30}{(m.)}     & \cellcolor{grey!30}{$L_2$} & \cellcolor{grey!30}{DCS}                      & \cellcolor{grey!30}{MM} &   \cellcolor{grey!30}{ICE}                         \\
           \hline
           \cellcolor{grey!30}{mean}   & \cellcolor{red!60}{0.44}  & 0.43 & \cellcolor{green!60}{0.41} & 0.42  \\
           \cellcolor{grey!30}{median}   & \cellcolor{red!60}{0.37}  & 0.36 & \cellcolor{green!60}{0.35} &\cellcolor{green!60}{0.35}  \\
           \cellcolor{grey!30}{std. dev.} & \cellcolor{red!60}{0.30}  & \cellcolor{green!60}{0.27} & 0.29 &0.28  \\
           \cellcolor{grey!30}{max}      & \cellcolor{red!60}{5.38} & 5.33 & 5.35 & \cellcolor{green!60}{5.22} \\
          \hline
      \end{tabular}%
      }
      \label{table:drive_1_hq}
 \end{subtable} \hfill
 \begin{subtable}{0.3\linewidth}
      \centering
      \caption{Localization results for data collect 2.}
      \resizebox{1.00\columnwidth}{!}{%
      \begin{tabular}{l|ccccc}
           \hline
           \cellcolor{grey!30}{(m.)}     & \cellcolor{grey!30}{$L_2$}                    & \cellcolor{grey!30}{DCS}                          & \cellcolor{grey!30}{MM}                        & \cellcolor{grey!30}{ICE}                       \\
           \hline
           \cellcolor{grey!30}{mean}   & \cellcolor{green!60}{0.79} & 0.81      & \cellcolor{red!60}{0.84}  & \cellcolor{green!60}{0.79} \\
           \cellcolor{grey!30}{median}   & 0.82 & \cellcolor{green!60}{0.81}      & \cellcolor{red!60}{0.84}  & 0.83 \\
           \cellcolor{grey!30}{std. dev.} & \cellcolor{green!60}{0.46}    & \cellcolor{green!60}{0.46} & \cellcolor{red!60}{0.50} & \cellcolor{green!60}{0.46}                     \\
           \cellcolor{grey!30}{max}    & 3.97  & 3.93  & \cellcolor{red!60}{10.77} &  \cellcolor{green!60}{2.95}                     \\
           \hline
      \end{tabular}%
      }
      \label{table:drive_2_hq}
 \end{subtable} \hfill
 \begin{subtable}{0.3\linewidth}
      \centering
      \caption{Localization results for data collect 3}
      \resizebox{1.0\columnwidth}{!}{%
          \begin{tabular}{l|ccccc}
           \hline
           \cellcolor{grey!30}{(m.)}     & \cellcolor{grey!30}{$L_2$}                     & \cellcolor{grey!30}{DCS}                       & \cellcolor{grey!30}{MM}                         & \cellcolor{grey!30}{ICE}                        \\
           \hline
          \cellcolor{grey!30}{mean}   & 1.09 & 1.10 & \cellcolor{red!60}{1.11}  & \cellcolor{green!60}{1.07} \\
          \cellcolor{grey!30}{median}   & 0.96 & 0.95 & \cellcolor{red!60}{1.00}  & \cellcolor{green!60}{0.89} \\
           \cellcolor{grey!30}{std. dev.} & \cellcolor{green!60}{0.67}  & \cellcolor{red!60}{0.73} & 0.72   & \cellcolor{green!60}{0.66}                      \\
           \cellcolor{grey!30}{max}      & 7.83  & 7.83  & \cellcolor{red!60}{18.08} & \cellcolor{green!60}{7.82}                      \\
       \hline
      \end{tabular}%
      }
      \label{table:drive_3_hq}
 \end{subtable}
 \label{table:hq_stats}
\end{table*}

\begin{figure*}[t]
    \begin{subfigure}{0.32\linewidth}
        \centering
        \includegraphics[width=0.95\linewidth]{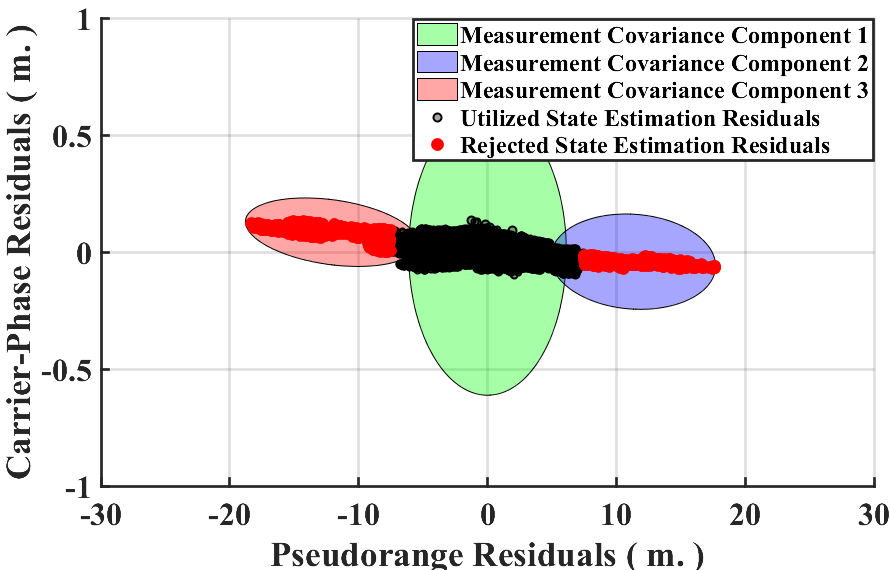} 
         \caption{Incrementally estimated measurement error covariance model for data collect 1. For this measurement uncertainty model, approximately 91\% of the observations are characterized by component 1.}
        \label{fig:inc_cov_ds1}
    \end{subfigure}\hfill
    \begin{subfigure}{0.305\linewidth}
        \centering
        \includegraphics[width=0.95\linewidth]{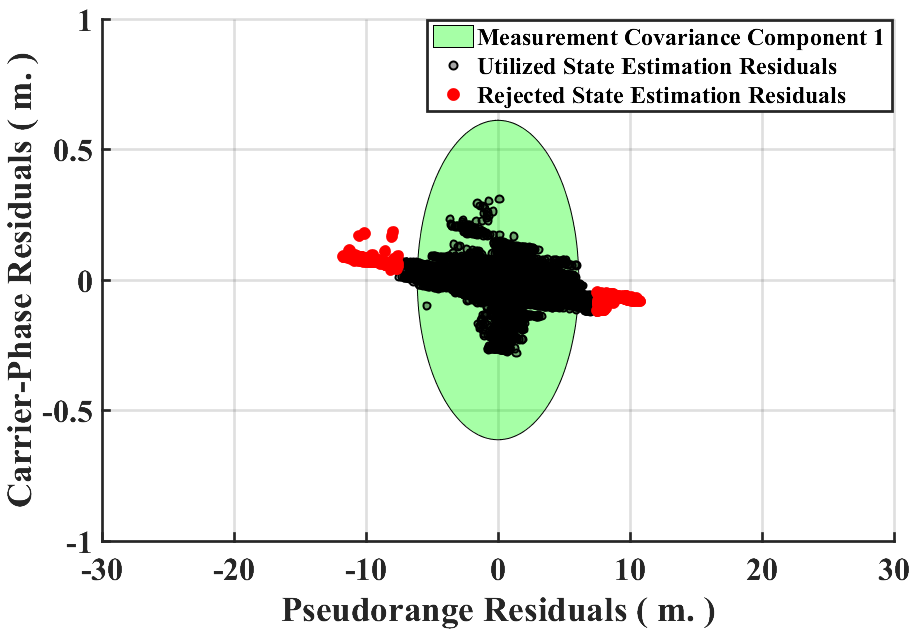}
        \caption{Incrementally estimated measurement error covariance model for data collect 2. For this data collect, only 249 observations did not adhere to the \apre measurement uncertainty model.}
\label{fig:inc_cov_ds2}
\end{subfigure} \hfill
    \begin{subfigure}{0.305\linewidth}
        \centering
        \includegraphics[width=0.95\linewidth]{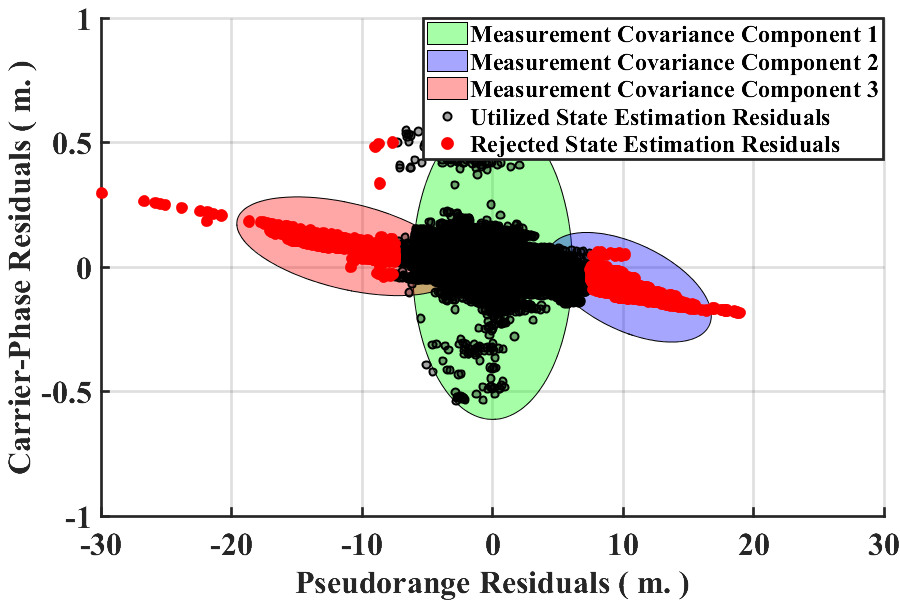}
        \caption{Incrementally estimated measurement error covariance model for data collect 3. For this measurement uncertainty model, approximately 98\% of the observations are characterized by component 1.}\label{fig:inc_cov_ds3}
\end{subfigure}

\caption{Incrementally estimated measurement error covariance model when the observations are generated with high fidelity receiver tracking parameters. }
    \label{fig:inc_cov}
\end{figure*}

To continue the localization performance evaluation, we can assess the localization performance of the four estimation frameworks with the high-quality observations, as provided in Table \ref{table:hq_stats}. From Table \ref{table:hq_stats}, first, it should be noted that all four estimation frameworks are providing comparable localization statistics -- as would be expected when the utilized observations adhere to the \apre measurement error covariance model. However, it can also be noted that the \ac{ICE} approach is providing the most accurate localization statistics the majority of the time.

\subsubsection{Covariance Estimation Analysis}

To continue the evaluation, the estimated covariance from the \ac{ICE} approach is assessed. Within this assessment, we have two primary objectives. First, we would like to show that the incrementally estimated covariance represents the measurement uncertainty model. Secondly, we would like to show that the covariance estimation process is efficiently conducted. 

To enable this assessment the high-quality observations are utilized, as provided in Fig. \ref{fig:inc_cov}. Within Fig. \ref{fig:inc_cov}, the black points correspond to the state estimation residuals of observations which sufficiently adhere to the \apre measurement error uncertainty model. While, the red points correspond to the state estimation residuals of observations which were not well defined by the \apre measurement uncertainty model, and thus not included during optimization; however, were utilized to modify the measurement uncertainty model. Additionally, the ellipses correspond to components of the incrementally estimated measurement error uncertainty model, with $95$\% confidence.

From Fig. \ref{fig:inc_cov}, it can be seen that the incrementally estimated measurement uncertainty models closely resemble the assumed model for the high quality observations (i.e., an inlier distribution which characterizes a majority of the observations, and outlier distributions which characterize a small percentage of erroneous observations). This is specifically evident for data collects 1 and 3, as depicted in Fig. \ref{fig:inc_cov_ds1} and Fig. \ref{fig:inc_cov_ds3}, respectively.



To verify the efficiency of the covariance adaptation approach, we can evaluate the number of times the measurement uncertainty model was adapted. For, data collects 1 and 3, as depicted in Fig, \ref{fig:inc_cov_ds1} and Fig. \ref{fig:inc_cov_ds3}, the covariance model was only adapted once to enable the incorporation of two outlier distributions. For data collect 2, as depicted in Fig. \ref{fig:inc_cov_ds2}, no covariance adaptation step was conducted -- instead, only $249$ observations were rejected. In contrast, if the covariance model was naively adapted every time the number of residuals were greater than the residual cardinality threshold\footnote{For this study, the threshold for measurement uncertainty model adaptation, was set to $1,000$ (i.e., adapt the uncertainty model if $|\mathbf{R_O}| > 1,000$).}, then data collect 1 would have required 75 adaptations, data collect 2 would have required 57 adaptations, and data collect 3 would have required 91 adaptations. Thus, the incorporation of the $\operatorname{z-test}$ to partition the set of residuals dramatically increased the efficiency of the proposed approach.

\subsubsection{Run-time Analysis}

\begin{figure}
 \centering
 \includegraphics[width=0.9\linewidth]{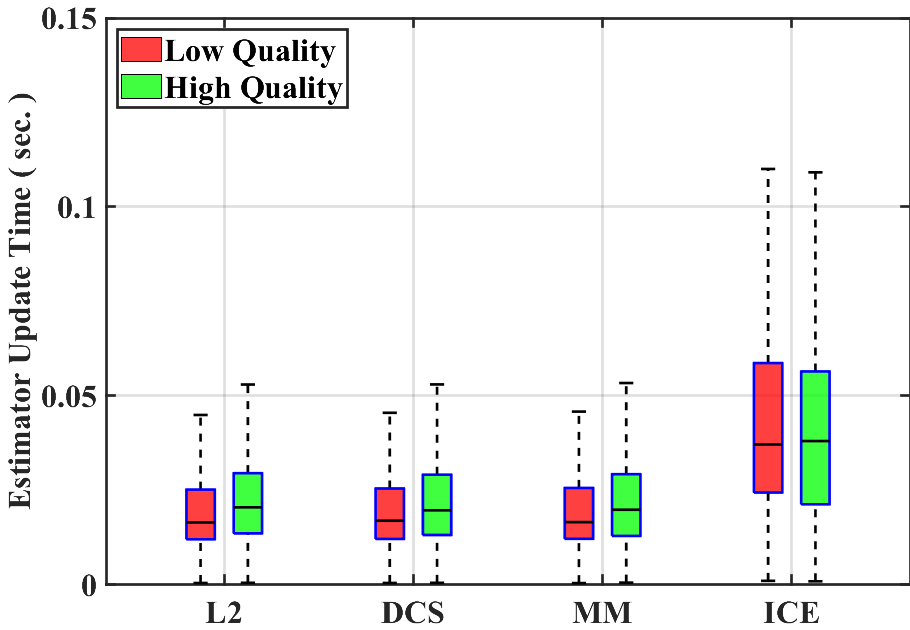}
 \caption{Estimator update time for each of the estimation frameworks over all data collects, where $L2$ is a batch estimator with $l^2\text{-norm}$ cost function, DCS is the dynamic covariance scaling robust estimator, MM is the max-mixtures approach with a static measurement covariance model, and ICE is the proposed incremental covariance estimation technique.}
 \label{fig:run_time}
\end{figure}

To conclude the evaluation of the proposed methodology, a run-time comparison\footnote{This run-time comparison was conducted on a 2.8GHz Intel Core i7-7700HQ processor.}  is provided in Fig \ref{fig:run_time}. From Fig. \ref{fig:run_time}, it is shown that $l^2\text{-norm}$, \ac{DCS}, and the \ac{MM} approaches all provide comparable run-time performance.

Additionally, it is clearly shown that the \ac{ICE} methodology, provides the slowest average run-time; however, this slower run-time -- which is still on average approximately $25 \ \text{Hz}$ -- could prove to be a valid comprise when considering the significantly increase in localization accuracy granted by the approach. 

Finally, although the \ac{ICE} approach does currently provide the slowest run-time, an additional points should be made. For the current \ac{ICE} implementation, the primary run-time bottle-neck for the current evaluation is implementation based. Specifically, the \ac{ICE} algorithm could be implemented in such a way to dramatically decrease run-time by simply parallelizing the covariance adaptation and state estimation steps.

    \section{Conclusion} \label{sec:conclusion}

Within this paper, we propose a novel extension to the measurement uncertainty model estimation paradigm for enabling robust state estimation. Specifically, we propose an efficient, incremental extension of the methodology. The efficiency of the approach is granted by adapting the uncertainty model with only a small subset of informative state estimation residuals (i.e., the state estimation residuals which do not adhere to the \apre model). The incremental nature of the approach is granted through recent advances within the probabilistics graphical model community, and the ability to merge \ac{GMM}'s.

To evaluate the proposed \ac{ICE} approach, three degraded GNSS data sets are utilized. Based upon the results obtained on these data sets, the proposed approach provides promising results. Specifically, the proposed \ac{ICE} approach provides significantly increased localization performance when utilizing degraded data, when compared to other state-of-the-art robust, incremental estimation algorithms.
    
    \bibliographystyle{ieeetr}
    \bibliography{main}

\end{document}